\documentstyle[12pt]{article}
\newcounter{mycount}

\newcommand{\be}{\begin{eqnarray}}
\newcommand{\ee}{\end{eqnarray}}
\newcommand{\bfl}{\begin{flushleft}}
\newcommand{\efl}{\end{flushleft}}
\newcommand\ie {{\it i.e. }}
\newcommand\eg {{\it e.g. }}

\newcommand\half{\frac 1 2 }

\newcommand\noi{\noindent}

\begin{document}

\bibliographystyle{nphys}

\centerline{\Large\bf Explicit 
Solution to the N-body Calogero problem}
\vspace* {-25 mm}
\begin{flushright} USITP-92-5 \\
May 1992
\end{flushright}

\vskip 0.9in
\centerline{L. Brink$^\circ$, T.H.\ Hansson$^\dagger$,  and
M. A. Vasiliev$^\star$ }

\vskip 2cm
\centerline{\bf ABSTRACT}
\vskip 3mm
We solve the N-body Calogero problem, \ie N particles in 1 dimension
subject to a two-body interaction of the form $\half \sum_{i,j}[ (x_i - x_j)^2
+
g/ {(x_i - x_j)^2}]$, by constructing annihilation and creation operators
of the form $ a_i^\mp =\frac 1 {\sqrt 2} (x  _i \pm i\hat{p}_i )$, where
$\hat{p}_i$ is a modified momentum operator obeying 
Heisenberg-type commutation relations with $x_i$, involving explicitly
permutation operators. On the other hand, $ D_j =i\,\hat{p}_j$ can be
interpreted as a covariant derivative corresponding to a flat connection.
The relation to fractional statistics in 1+1 dimensions and anyons in a
strong magnetic field is briefly discussed.

\vfill\noi
$^\circ$Institute of Theoretical  Physics, CTH, S-41296 G{\"o}teborg, Sweden.
\vskip 2mm\noi
$^\dagger$Institute of Theoretical Physics, University of Stockholm,
Vanadisv\"agen 9, \\
S-113 46 Stockholm, Sweden.
\vskip 2mm\noi
$^\star$Department of Theoretical Physics, P. N. Lebedev Physical Institute,
\\
117924 Leninsky Prospect 53, Moscow, Russia.

\vskip 3mm \noi
$^{\circ\, \dagger}$Supported by the Swedish Natural Science Research Council

\eject

In this letter we present an operator solution to the Calogero problem, \ie we
find the eigenvalues and eigenfunctions of the Hamiltonian
\be
H_{Cal} =\half \sum_{i=1}^N \left[ -\partial_i^2 + x_i^2 \right] +
\sum_{j < i}^N \frac g {(x_i-x_j)^2} \ \ \ \ \ ,
\ee
which is known to be completely integrable. The eigenvalues were
found by Calogero\cite{calo1,calo2} who also found the wave functions for
$N=3$ and $N=4$. \footnote
{ Our Hamiltonian differs from that in
reference \cite{calo2} by an overall normalization and an
inessential harmonic oscillator term for the center of mass coordinate.}
Some of these results were also obtained by Sutherland\cite{suth12}, and
later the $N=5$ wave functions were constructed by Gambardella\cite{gamb1}.
For a comprehensive review see\cite{olsh1,olsh2}.
The main new result in this paper are explicit
expressions for the N body wavefunctions, but our derivation is also
considerably simpler than the original one and emphasizes the interpretation
in terms of fractional statistics\cite{lein3,polyc1,hans6}.

To solve $H_{Cal} \Psi = E\Psi$, we write
\be
\Psi^\pm = \prod_{i>j}
(x_i -x_j)^{\nu} \Phi^\pm = \beta^{\nu}\Phi^\pm\ee
where
$x_i >x_j $ for $i\,>\,j$, while $+$ and $-$  refers to totally symmetric and
antisymmetric wavefunctions $\Phi^+$ and $\Phi^-$, respectively, 
and define a covariant derivative by
\be
 D_i =\partial_i + \nu \sum_{j\neq i} (x_i -x_j )^{-1} (1-K_{ij} )
\ee
where $K_{ij}$ is the permutation operator which interchanges the
particles with labels $i$ and $j$. Essentially the same construction was found
by Polychronakos, who in a recent paper used covariant derivatives to derive
the constants of motion for the Calogero problem\cite{polyc3}.
After some algebra we find
\be
 H_{Cal}\Psi^\pm = \beta^{\nu} \half (-D^2 + X^2) \Phi^\pm
\ee
where $D^2 = \sum_{i=1}^N D_i^2$, $X^2 = \sum_{i=1}^N x_i^2$ and
$g=\nu(\nu\mp 1 )$, where the upper and lower sign refers to symmetric and
antisymmetric wave functions respectively.
By direct calculation, we also find the commutation
relations,
\be
[D_i ,x_j] =A_{ij}=\delta_{ij} (1+\nu \sum_{k=1}^N K_{ik} )-\nu K_{ij}
\ee
and
\be
 [D_i ,D_j] =0 \ \ \ \ \ ,
\ee
results that were also found in \cite{polyc3}.
In deriving these results, we used both  defining
relations for the permutation operator like, $K_{ij}K_{ik}=K_{jk}K_{ij}$, and
commutation relations like $K_{ij} x_j =x_i K_{ij}$. Note that the commutator
(6) is remarkably simple considering the complicated form of $D_i$, and that
$A_{ij}$ in (5) is symmetric in $i$ and $j$ so that we can construct creation
and annihilation operators via
\be
 a_i^\mp =\frac 1 {\sqrt 2} (x	_i \pm D_i )
\ee
obeying the commutation relations
\be
 [a_i^\pm ,a_j^\pm] &=& 0
\ee
\be
[a_i^- ,a_j^+] &=& A_{ij} \ \ \ \ \ .
\ee
The Hamiltonian can now be expressed as
\be
H= \half (-D^2 + X^2) =\frac{1}{2}\sum_i \{a_i^+ ,a_i^- \} \ \ \ \ \ ,
\ee
and turns out to obey the standard commutation relations with the creation
and annihilation operators,
\be
[H,a_i^\pm ] = \pm a_i^\pm \ \ \ \ \ .
\ee
This last relation again follows from a series of nontrivial algebraic
manipulations using the properties of the $K_{ij}$.

The eigenfunctions are now obtained via the standard construction, \ie
\be
\Phi^\pm(n_i) =  {\cal {S_\pm}}\{ \prod_{i=1}^N (a_i^+)^{n_i}\} \Phi_0^\pm
      			\ \ \ \ \ ,
\ee
where ${\cal S _\pm}$ denotes total (anti)symmetrization, and the vacuum state
$\Phi_0^\pm$ satisfies
 \be
 a^-_i \Phi_0^\pm = 0 \ \ \ \ \ ,
\ee
and
\be
 K_{ij}\Phi_0^\pm =\pm \Phi_0^\pm \ \ \ \ \ .
\ee
Using this and the commutation relations (11) we find the ground state
energy of $H_{Cal}$ to be $E_0^\pm = \frac N 2 \pm \nu\half N(N-1) $,
so the complete spectrum is that of N
bosons or fermions in a harmonic oscillator well shifted by this constant.
This is Calogero's original result. Solving (13) and (14) also immediately
gives Calogero's ground state wavefunction.
As advertised, the new result is the explicit expression (12) for the
wavefunctions. Needless to say, the expressions very quickly become very
cumbersome because of the sums in the definition of $D_i$.

We now make some comments to the above:

1. If in the case N=2 we separate into relative and CM coordinates, the
Calogero Hamiltonian for the relative coordinate, $x= x_1 - x_2$, takes the
form $H_{rel} = -\partial^2 + x^2/4 +g/x^2$, and the relevant
raising and lowering
operators read $a^\pm =\frac x 2 \mp \partial \mp \frac \nu x (1-K)$, $K$
anticommuting with $x$ and $\partial$. 
It is easy to check that the operators $A= \half \{ a^- , a^+ \}$ and $B_\pm
= \half (a^\pm)^2$, generate an sp(1,R) algebra. The quadratic Casimir
invariant is given by $\Gamma = [-3/4 - \nu K(1 - \nu K)]/4$, thus reducing
to $\Gamma = [-3/4 + \nu(\nu - 1)]/4$ on symmetric wavefunctions. 
This should be compared with the standard group-theoretical solution of the
$N=2$ problem\cite{pere2,lein3} where the generators are given by
$A= \half \{a_0^- , a^+_0\}+ g/x^2$ 
and $B_\pm = \half [(a^\pm_0)^2 - g/x^2] $
(with $a^\pm_0 = \frac x 2 \mp \partial$), and the corresponding Casimir
invariant  $\Gamma = [-3/4 + g]/4$, so we again obtain $g = \nu(\nu -1)$.
That the sp(1,R) generators could be expressed as bilinears of creation and
annihilation operators obeying the modified commutation relations involving the
Klein-type operator $K$ even for $\Gamma \neq -3/16$ was noted in \cite{vasi1}
in relation with higher-spin gauge theories in four dimensions, 
and it was this observation that triggered the present investigation.

2. The connection between the Calogero problem and fractional statistics
in 1+1 dimension was first noted in \cite{lein3}, and the N-body problem was
discussed in \cite{polyc1,hans6}.
Equation (6) is suggestive of a flat connection on the configuration space 
$R^N /S_N $ with $x_i > x_j $ for $i>j$, where $S_N$ denotes the symmetric 
group. Because of the permutation operators 
in the $D_i$:s we must consider these as acting on $N\,!$ dimensional vectors,
where the components are the values of $\Phi (x_i)$ at points connected by
permutation of the coordinates, \ie $\Phi(x_i) \rightarrow \Phi^\sigma (x_i)
= \Phi (\sigma[x_i])$, where $\sigma$ is a permutation. The covariant
derivative can now be written on matrix form \be
D^{\sigma,\sigma '}_i = \delta^{\sigma,\sigma '} \partial_i + \nu
   \sum_{j\neq i}^N \frac 1 {x_i - x_j} (\delta^{\sigma,\sigma '}
      - \delta^{\sigma,\tau_{ij}\sigma '}{\cal K}_{ij})
\ee
where $\tau_{ij}$ is the transposition of $i$ and $j$, and where the ${\cal
K}_{ij}$:s fulfill the same commutation relations with $x_i$, $\partial_i$
and themselves, as the $K_{ij}$:s, but do not affect the wave functions.
Because of the integrability
condition (6), one would normally be able to, at least locally, find a $g$
so that $g^{-1}\partial_i g = D_i$. In our case this construction might be
obstructed, since the would be potential $A_i = g^{-1}(\partial_i g)$ contains
permutation operators, and thus $g$ would not in general commute with
$\partial_i$. This is in
fact satisfying, since if a (global) $g$ could be found, then
the spectrum of $H_{Cal}$ would be identical to
that of the harmonic oscillator.
We believe that a more in depth study of the geometrical aspects of the
$N$-body problem could be rewarding, and hope to return to that in future work.

3. By analogy with the standard Heisenberg commutation relations, one can take
(5) and (6) as the defining relations for a generalized algebra.
{}From this viewpoint, what is remarkable is
that these commutation relations are consistent, \ie they satisfy
the Jacobi identities.
In the coordinate representation where the coordinate operator $\hat x_i$ acts
by multiplication with $x_i$, one can then derive the explicit form (3) for
the momenta $\hat{p}_j =-iD_j$. Such an interpretation might \eg be useful
in order to define an appropriate
momentum representation (the definition of a  generalized Fourier transform is
in this case not at all obvious.)  Let us emphasize that since the
generalized commutation
relations (5) and (6) explicitly involve the permutation operators, they
account
in a rather nontrivial way for effects due to identical particles, and are
adequate for describing fractional statistics.

4. There is a close similarity between the
Calogero problem and N anyons in the lowest Landau
level\cite{hans6,polyc2}.
In \cite{hans6} it was shown that the system of
two anyons in the lowest Landau level is in fact equivalent to the 2-body
Calogero problem. It was also shown that, after appropriate rescalings, the
spectrum of the total angular
momentum operator for N anyons is identical to that of the N-body Calogero
Hamiltonian. The wave functions for anyons in the lowest Landau level are
all explicitly known, and can in fact be constructed from with the help of
raising and lowering operators. Our operator construction of the
Calogero wave functions strongly supports the conjecture in \cite{hans6} that
the systems are in fact equivalent. We are presently trying to find an
explicit mapping between the wave functions.

\vskip 3mm\noi
{\bf Acknowledgements} \\
H. Hansson thanks T. Ekedahl and M. Vasiliev thanks O. V. Ogievetsky  and
M. A. Olshanetsky for useful discussions.

\newpage

\vfill
\end{document}